\documentclass[conference]{IEEEtran}
\IEEEoverridecommandlockouts
\usepackage{cite}
\usepackage{amsmath,amssymb,amsfonts}
\usepackage{algorithmic}
\usepackage{url}
\usepackage{graphicx}
\usepackage{textcomp}
\usepackage{epstopdf}
\usepackage{xcolor}
\usepackage{placeins}
\def\BibTeX{{\rm B\kern-.05em{\sc i\kern-.025em b}\kern-.08em
    T\kern-.1667em\lower.7ex\hbox{E}\kern-.125emX}}
\begin{document}

\title{Deepfake Audio Detection Using Self-supervised  Fusion Representations}

\author{
\IEEEauthorblockN{Khalid Zaman, Qixuan Huang, Muhammad Uzair, Masashi Unoki}
\IEEEauthorblockA{
Graduate School of Advanced Science and Technology\\
Japan Advanced Institute of Science and Technology\\
1-1 Asahidai, Nomi, Ishikawa, Japan 923-1292\\
%
}
}

\maketitle

\begin{abstract}

This paper describes a submission to the Environment-Aware Speech and Sound Deepfake Detection Challenge (ESDD2) 2026, which addresses component-level deepfake detection using the CompSpoofV2 dataset, where speech and environmental sounds may be independently manipulated. To address this challenge, a dual-branch deepfake detection framework is proposed to jointly model speech and environmental contextual representations from input audio. Two pretrained models, XLS-R for speech and BEATs for environmental sound, are used to extract complementary contextual representations. A Matching Head is introduced to model representation differences through statistical normalization and representation interaction, enabling estimation of the original class. In parallel, multi-head cross-attention enables effective information exchange between speech and environmental components. The refined representations are processed with residual connections and layer normalization, and passed to an AASIST classifier to predict speech-based and environment-based spoofing probabilities. The model outputs original, speech, and environment predictions. On the test set, the proposed system achieves an F1-score of 70.20\% and an environmental EER of 16.54\%, outperforming the baseline system.

\end{abstract}

\begin{IEEEkeywords}
Self-supervised learning, Wav2vec 2.0, BEATs, deepfake detection, and deep learning.
\end{IEEEkeywords}

\section{Introduction}
\label{sec:intro}

Recent advances in speech generation technologies, such as text-to-speech and voice conversion, have significantly improved the quality and naturalness of synthesized audio\cite{shen2018natural, kim2021conditional, li2023styletts, casanova2022yourtts, li2021starganv2}. While these developments enable many beneficial applications, they also raise serious concerns regarding audio deepfakes, where manipulated or synthesized audio can be used to impersonate individuals or spread misleading information. Consequently, detecting spoofed or manipulated audio has become an important research topic in the field of audio forensics\cite{todisco2019asvspoof, jung2025spoofceleb, alali2025partial}.

A wide range of speech spoofing detection methods and benchmark datasets have been proposed in the literature. Several large-scale initiatives, including the ASVspoof Challenge \cite{todisco2019asvspoof, delgado2021asvspoof}, and the ADD Challenge \cite{yi2022add}, along with widely used datasets such as SpoofCeleb\cite{jung2025spoofceleb} and WaveFake\cite{frank2021wavefake}, have provided standardized evaluation protocols and diverse data sources for detecting synthesized, converted, and replayed speech. Early approaches relied on handcrafted acoustic representations such as Mel-frequency cepstral coefficients (MFCCs), constant-Q cepstral coefficients (CQCCs), and spectrogram-based representations, combined with traditional classifiers including Gaussian mixture models (GMMs), convolutional neural networks (CNNs)\textbf{,} and support vector machines (SVMs) \cite{hamza2022deepfake, zaman2024hybrid, nautsch2021asvspoof, yang2026deep}. Although these methods demonstrated promising results, they often struggled to generalize to unseen spoofing attacks and diverse recording conditions.

More recently, self-supervised learning (SSL) models have been widely adopted for deepfake speech and audio detection. SSL-based approaches, such as Wav2vec 2.0, HuBERT, WavLM, and BEATs, learn rich contextual and semantic representations from large-scale unlabeled audio data and have demonstrated strong robustness and generalization performance \cite{wang2025mixture,li2023voice,combei2024wavlm, pianese2024training}.


However, most existing methods rely on a single SSL model and process the audio signal as a whole, without explicitly considering the distinct representations of its underlying components. In realistic scenarios, speech and environmental sounds exhibit different properties and may be manipulated independently, leading to latent inconsistencies within the audio. Consequently, treating the input as a unified signal limits the ability of current approaches to capture such component-level variations and interactions.

Therefore, this study proposes a component-level deepfake detection framework based on dual SSL representations\footnote{\url{https://github.com/OrgHuang/KHUM-ESDD2.git}}. Specifically, two pretrained models, XLS-R \cite{babu2022xls} for speech and BEATs \cite{chen2023beats} for environmental sound, are used to extract complementary contextual representations. A Matching Head is introduced to model the discrepancies between XLS-R and BEATs representations through statistical normalization and representation interaction, enabling estimation of the original class probability. In parallel, a dual-branch architecture with multi-head cross-attention facilitates effective information exchange between speech and environmental components. The refined representations are further processed using residual connections and layer normalization, and then passed to an AASIST classifier \cite{jung2022aasist} to independently predict speech-based and environment-based spoofing probabilities. This design enables the model to better capture inconsistencies between manipulated and genuine components, improving robustness against environment-aware spoofed audio.

The rest of the paper is organized as follows. Section 2 describes the dataset in detail. Section 3 presents the proposed method. Section 4 reports the experimental setup, results, and discusses the findings. Finally, Section 5 concludes the paper.

\begin{table}[t]
\centering
\caption{Statistical overview of the CompSpoofV2 dataset.}
\label{tab:dataset_stats}
\resizebox{\linewidth}{!}{%
\begin{tabular}{lcccc}
\hline
\textbf{Class} & \multicolumn{4}{c}{\textbf{Dataset Split}} \\ \cline{2-5}
 & \textbf{Train} & \textbf{Val} & \textbf{Eval} & \textbf{Test} \\
\hline
Original & 48,639 & 6,939 & 7,455 & 7,415 \\
Bonafide\_bonafide & 25,189 & 2,784 & 3,570 & 3,635 \\
Spoof\_bonafide & 21,759 & 2,413 & 2,980 & 2,987 \\
Bonafide\_spoof & 50,361 & 8,071 & 7,655 & 7,672 \\
Spoof\_spoof & 29,413 & 4,657 & 5,945 & 5,894 \\
\hline
Total & 175,361 & 24,864 & 27,605 & 27,603 \\
\hline
\end{tabular}%
}
\end{table}

\setlength{\textfloatsep}{6pt}
\setlength{\dbltextfloatsep}{6pt}

\begin{figure*}[htbp]
\centering
\includegraphics[width=.92 \linewidth]{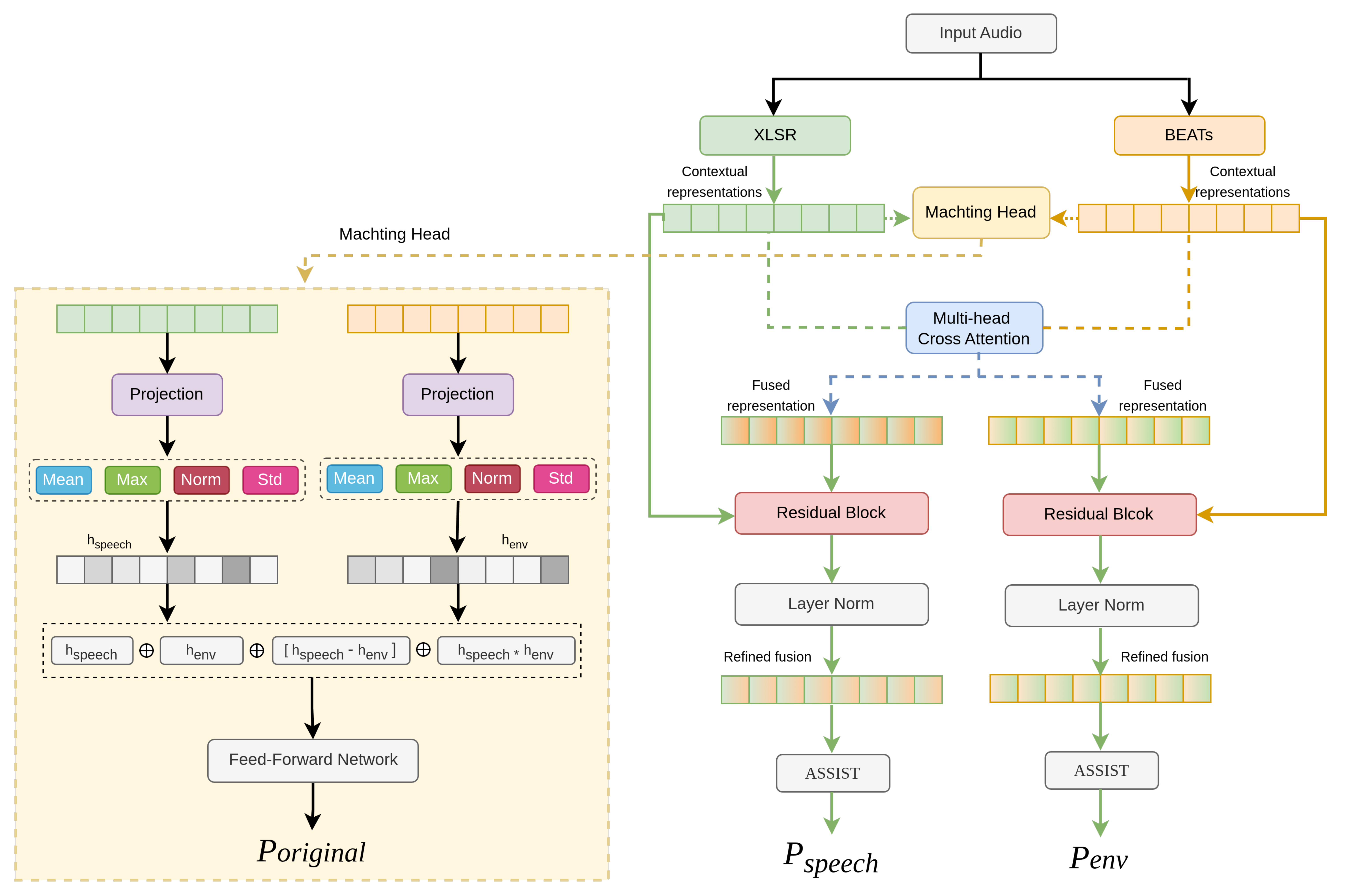} 
\caption{Proposed dual-branch model for component-level spoofing detection.}
\label{pm}
\end{figure*}

\section{Dataset}

All experiments are conducted on the CompSpoofV2 dataset, introduced for the ESDD2 challenge on component-level speech and environmental sound spoofing detection~\cite{zhang2026esdd2}. The dataset contains over 250,000 audio samples with a total duration of approximately 283 hours. Each sample has a fixed duration of 4 seconds and is provided at multiple sampling rates to simulate real-world acoustic and system-level variations. The CompSpoofV2 dataset is designed for component-level manipulation, where speech and environmental sound components can be independently genuine or spoofed. Each audio sample belongs to one of five classes, covering all combinations of bona fide and spoofed speech and environmental sound: original, bonafide\_bonafide, spoof\_bonafide, bonafide\_spoof, and spoof\_spoof. The dataset is divided into training, validation, evaluation, and test partitions, as summarized in Table~\ref{tab:dataset_stats}. The training and validation sets share the same data sources and class distributions. Similarly, the evaluation and test sets follow the same data sources and class distributions; however, they include newly generated audio samples that are unseen during training and validation.

\section{Proposed Method}

\subsection{Data Augmentation}

To improve the robustness of the model and better simulate diverse real-world conditions, we design a task-aware data augmentation approach that constructs various combinations of speech and environmental components.

We generate augmented audio samples using multiple mixing methods. Specifically, we consider several types of combinations, including original mixed audio, concatenation-based mixing, weighted summation, partial mixing, and time-shifted mixing. These methods aim to simulate different levels of consistency and mismatch between speech and environmental components.

Formally, given speech $s$ and environmental sound $e$, the augmented sample $x$ is constructed as:
\begin{equation}
x = \mathcal{A}(s, e; \theta),
\end{equation}
where $\mathcal{A}(\cdot)$ denotes the augmentation function and $\theta$ represents randomly sampled parameters such as the mixing ratio, temporal offset, and segment boundaries.

In addition, we introduce stochastic noise perturbation by injecting Gaussian noise with a randomly sampled signal-to-noise ratio (SNR), which further enhances robustness to acoustic degradation.

To address class imbalance, we adopt a class-aware sampling scheme during training. Minority classes are oversampled by repeating samples with different augmentation instances, ensuring a more balanced training distribution while maintaining diversity.

This augmentation framework enables the model to learn discriminative representations under varying degrees of correlation between speech and environmental components.

\subsection{Model Architecture}
The proposed model architecture consists of two parallel pretrained encoders, a bidirectional cross-attention module, two AASIST classifiers for speech and environmental spoofing detection, and a Matching Head for original class estimation, as illustrated in Fig.~\ref{pm}.

The model processes each input audio signal using two parallel pretrained encoders. XLS-R is used to extract speech-related representations, while BEATs is used to capture environmental representations. Given an input audio $x$, the two encoders produce:
\begin{equation}
F_\mathrm{speech} = \mathrm{XLS\text{-}R}(x), \quad 
F_\mathrm{env} = \mathrm{BEATs}(x),
\end{equation}
where $F_\mathrm{speech}$ and $F_\mathrm{env}$ denote the speech and environmental representations, respectively.

To perform speech and environmental spoofing detection, a bidirectional cross-attention module is employed to enable interaction between the two representation streams. Since the output dimensions of XLS-R and BEATs differ, the speech representations are first projected to match the dimensionality of the environmental representations:
\begin{equation}
\hat{F}_\mathrm{speech} = \mathrm{Linear}(F_\mathrm{speech}).
\end{equation}
where $\hat{F}_\mathrm{speech}$ denotes the projected speech representation aligned to the dimensionality of $F_\mathrm{env}$.

The bidirectional cross-attention interaction is formulated as:
\begin{equation}
\begin{aligned}
F_\mathrm{speech}^{*} &= \mathrm{CrossAttn}(\hat{F}_\mathrm{speech}, F_\mathrm{env}), \\
F_\mathrm{env}^{*} &= \mathrm{CrossAttn}(F_\mathrm{env}, \hat{F}_\mathrm{speech}),
\end{aligned}
\end{equation}
where $F_\mathrm{speech}^{*}$ and $F_\mathrm{env}^{*}$ represent the refined speech and environmental representations.

These refined representations are passed to two AASIST classifiers:
\begin{equation}
\begin{aligned}
P_{\mathrm{speech}} &= \mathrm{AASIST}_\mathrm{speech}(F_\mathrm{speech}^{*}), \\
P_{\mathrm{env}} &= \mathrm{AASIST}_\mathrm{env}(F_\mathrm{env}^{*}). 
\end{aligned}
\end{equation}
where $P_{\mathrm{speech}}$ and $P_{\mathrm{env}}$ denote the spoofing prediction logits for the speech and environmental branches, respectively.

For original class estimation, a Matching Head is introduced to model representation differences between speech and environmental streams. This module operates on the original encoder outputs prior to cross-attention. First, both representations are projected into a shared latent space:
\begin{equation}
H_\mathrm{speech} = \phi_\mathrm{speech}(F_\mathrm{speech}), \quad
H_\mathrm{env} = \phi_\mathrm{env}(F_\mathrm{env}),
\end{equation}
where $\phi_\mathrm{speech}(\cdot)$ and $\phi_\mathrm{env}(\cdot)$ denote linear projection layers, and $H_\mathrm{speech}$ and $H_\mathrm{env}$ denote the projected representations.

Next, statistical pooling is applied along the temporal dimension, and the resulting statistics are concatenated representations:
\begin{equation}
\begin{aligned}
h_\mathrm{speech} = \mathrm{Concat}(&\mathrm{Mean}(H_\mathrm{speech}), \mathrm{Max}(H_\mathrm{speech}), \\
                        &\mathrm{Std}(H_\mathrm{speech}), \mathrm{Norm}(H_\mathrm{speech}))
\end{aligned}
\end{equation}

\begin{equation}
\begin{aligned}
h_\mathrm{env} = \mathrm{Concat}(&\mathrm{Mean}(H_\mathrm{env}), \mathrm{Max}(H_\mathrm{env}), \\
                        &\mathrm{Std}(H_\mathrm{env}), \mathrm{Norm}(H_\mathrm{env}))
\end{aligned}
\end{equation}
where $\mathrm{Mean}(\cdot)$, $\mathrm{Max}(\cdot)$, and $\mathrm{Std}(\cdot)$ denote temporal average, maximum, and standard deviation pooling, respectively, and $\mathrm{Norm}(\cdot)$ denotes the $\ell_2$ norm computed along the temporal dimension and $h_\mathrm{speech}$ and $h_\mathrm{env}$ denote the concatenated representations.

The pooled features are combined using interaction operations:
\begin{equation}
\begin{aligned}
z = \mathrm{Concat}(&h_\mathrm{speech}, h_\mathrm{env}, |h_\mathrm{speech} - h_\mathrm{env}|, \\
                    &h_\mathrm{speech} \odot h_\mathrm{env})
\end{aligned}
\end{equation}
where $\odot$ denotes element-wise multiplication, and $z$ denotes the combined feature representation for original class estimation.

Finally, the combined feature vector is passed through a Feed-Forward Network (FFN):
\begin{equation}
P_{\mathrm{original}} = \mathrm{FFN}(z).
\end{equation}
where $P_{\mathrm{original}}$ denotes the output logits for original class prediction.

The model jointly produces three outputs:
\begin{equation}
(P_{\mathrm{speech}}, P_{\mathrm{env}}, P_{\mathrm{original}}),
\end{equation}
corresponding to speech spoofing detection, environmental spoofing detection, and original class estimation.


\begin{table}[t]
\centering
\caption{Performance of baseline and proposed methods across validation, evaluation, and test sets.}
\label{tab:baseline_results}
\resizebox{\columnwidth}{!}{%
\begin{tabular}{llcccc}
\hline
\textbf{Method} & \textbf{Dataset} & \multicolumn{4}{c}{\textbf{Evaluation Metrics (\%)}} \\ \cline{3-6}
 & & \textbf{F1-score} & \textbf{Original EER } & \textbf{Speech EER } & \textbf{Env. EER } \\
\hline
Baseline & Validation Set & 94.62 & 0.31 & 1.72 & 37.66 \\
Proposed & Validation Set & 94.37 & 0.63 & 6.76 & 10.64 \\
Baseline & Evaluation Set & 62.24 & 1.74 & 19.93 & 43.36 \\
Proposed & Evaluation Set & 70.11 & 2.99 & 31.40 & 16.54 \\
Baseline & Test Set & 63.27 & 1.73 & 19.78 & 42.79 \\
Proposed & Test Set & 70.20 & 2.59 & 32.98 & 18.83 \\
\hline
\end{tabular}%
}
\end{table}


\begin{table*}[t]
\begin{center}
\caption{Performance comparison and ablation study on the evaluation set. The results analyze the contribution of different components, including sampling strategy, data augmentation, loss weighting, and cross-attention-based fusion. ``w/o'' denotes removing the corresponding module, and ``Frozen Encoders'' indicates that the pretrained encoders are kept fixed. In the proposed model, only the last two layers of the encoders are fine-tuned.}

\label{tab3}
\resizebox{\textwidth}{!}{%
\begin{tabular}{l c c cc c cccc}
\hline
\textbf{Model} & \textbf{Time} & \textbf{Loss} 
& \multicolumn{2}{c}{\textbf{Data Augmentation}} 
& \textbf{Fusion} 
& \multicolumn{4}{c}{\textbf{Evaluation Metrics (\%)}} \\

\cline{4-5} \cline{7-10}

 &  &  & \textbf{Sampling} & \textbf{Aug. Type} 
 &  & \textbf{F1-score} & \textbf{Original EER} & \textbf{Speech EER} & \textbf{Env. EER} \\

\hline

Baseline & 4 hours & Avg. & None & None & No & 62.24 & \textbf{1.74} & \textbf{19.93} & 43.36 \\

Light Transformer & 3 hours & Avg. & None & None & No & 63.77 & 4.02 & 24.97 & 38.08 \\

XLS-R & 33 mins & Avg. & None & None & No & 61.81 & 3.82 & 32.15 & 30.60 \\

BEATs & 10 mins & Avg. & None & None & No & 62.11 & 4.71 & 27.67 & 33.36 \\

\hline

BEATs & 6 mins & Avg. & Minority & None & No & 62.74 & 5.41 & 32.12 & 31.67 \\

BEATs & 18 mins & Avg. & Majority & None & No & 64.95 & 4.98 & 32.26 & 26.83 \\

BEATs & 18 mins & Task-weighted & Majority & Mix & No & 63.56 & 4.58 & 28.65 & 24.79 \\

\hline

w/o Matching Head & 1 hour 11 mins & Task-weighted & Majority & Mix & Cross-attn & 66.07 & 4.01 & 29.43 & 22.74 \\

Frozen Encoder & 53 mins & Task-weighted & Majority & Mix & Cross-attn & 65.91 & 2.87 & 31.39 & 22.68 \\

Env w/o Fusion & 53 mins & Task-weighted & Majority & Mix & Cross-attn & 67.18 & 2.82 & 27.91 & 19.23 \\

Proposed & 1 hour & Task-weighted & Majority & Mix & Cross-attn & \textbf{70.11} & 2.99 & 31.40 & \textbf{16.54} \\

\hline
\end{tabular}%
}
\end{center}
\end{table*}


\setlength{\textfloatsep}{6pt}
\setlength{\dbltextfloatsep}{6pt}
\begin{figure*}[!t]
\centering
\setlength{\abovecaptionskip}{3pt}
\setlength{\belowcaptionskip}{0pt}

\begin{minipage}[b]{0.30\textwidth}
  \centering
  \includegraphics[width=\linewidth]{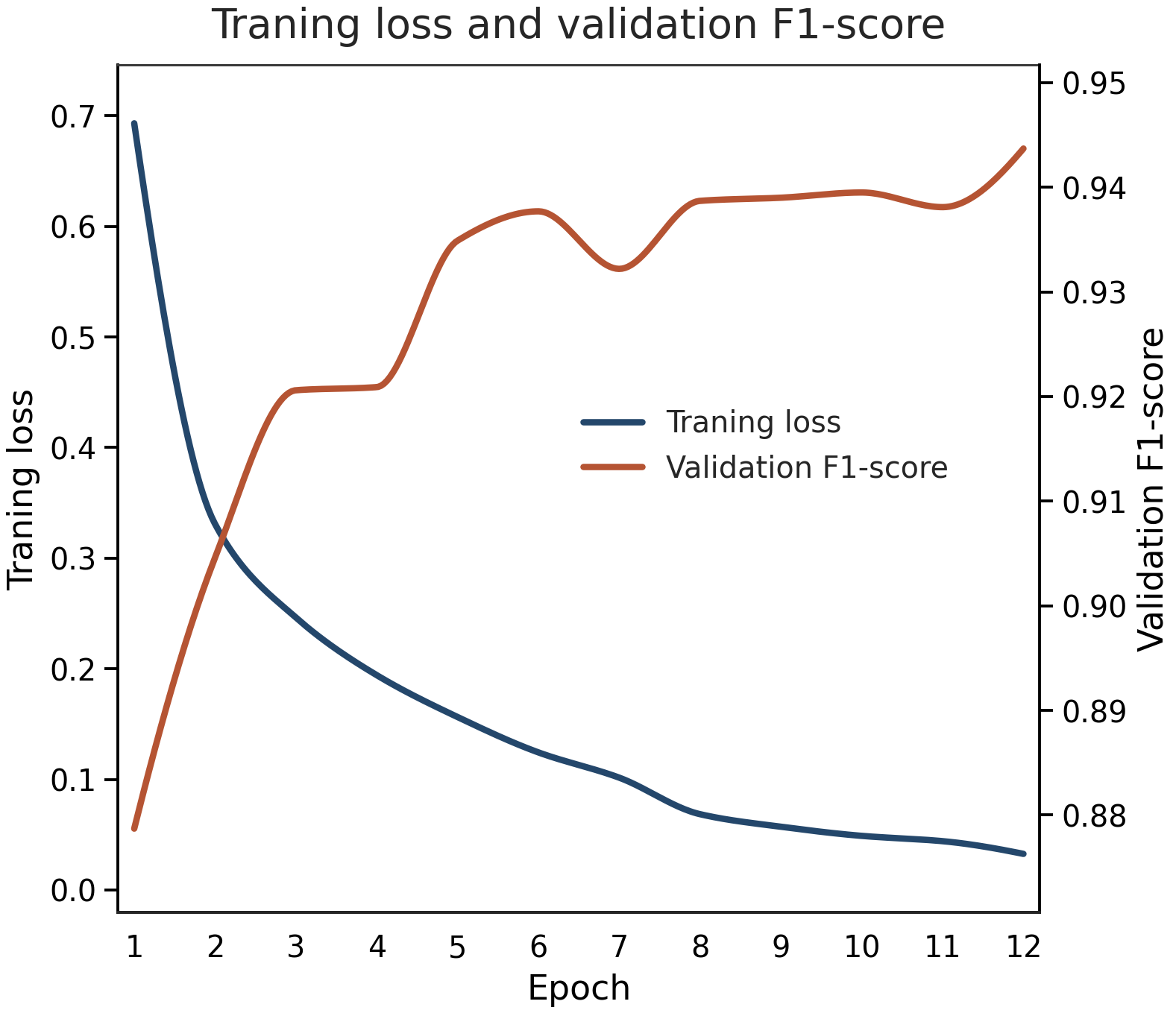}
  {\small (a)}
\end{minipage}
\hfill
\begin{minipage}[b]{0.32\textwidth}
  \centering
  \includegraphics[width=\linewidth]{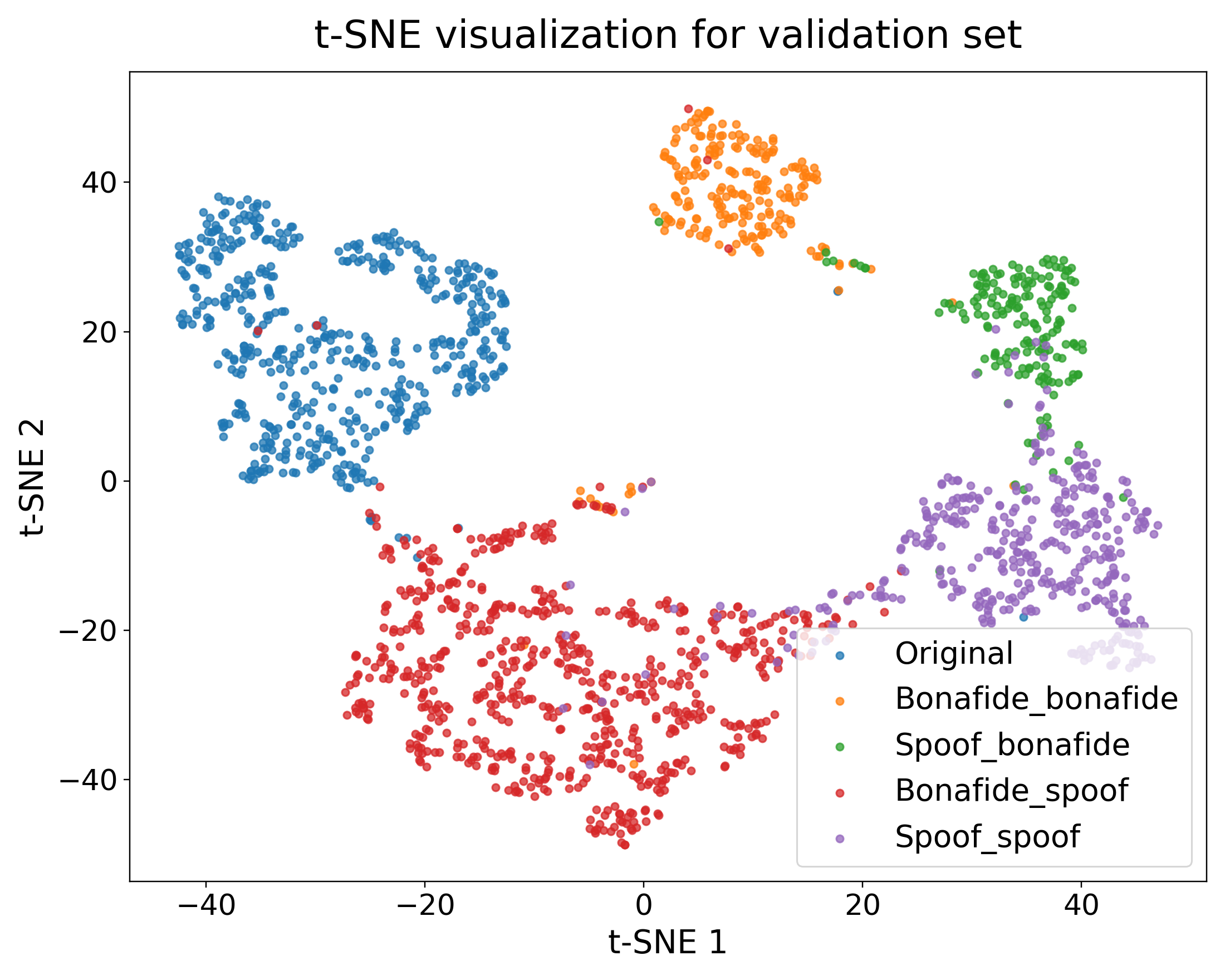}
  {\small (b)}
\end{minipage}
\hfill
\begin{minipage}[b]{0.33\textwidth}
  \centering
  \includegraphics[width=\linewidth]{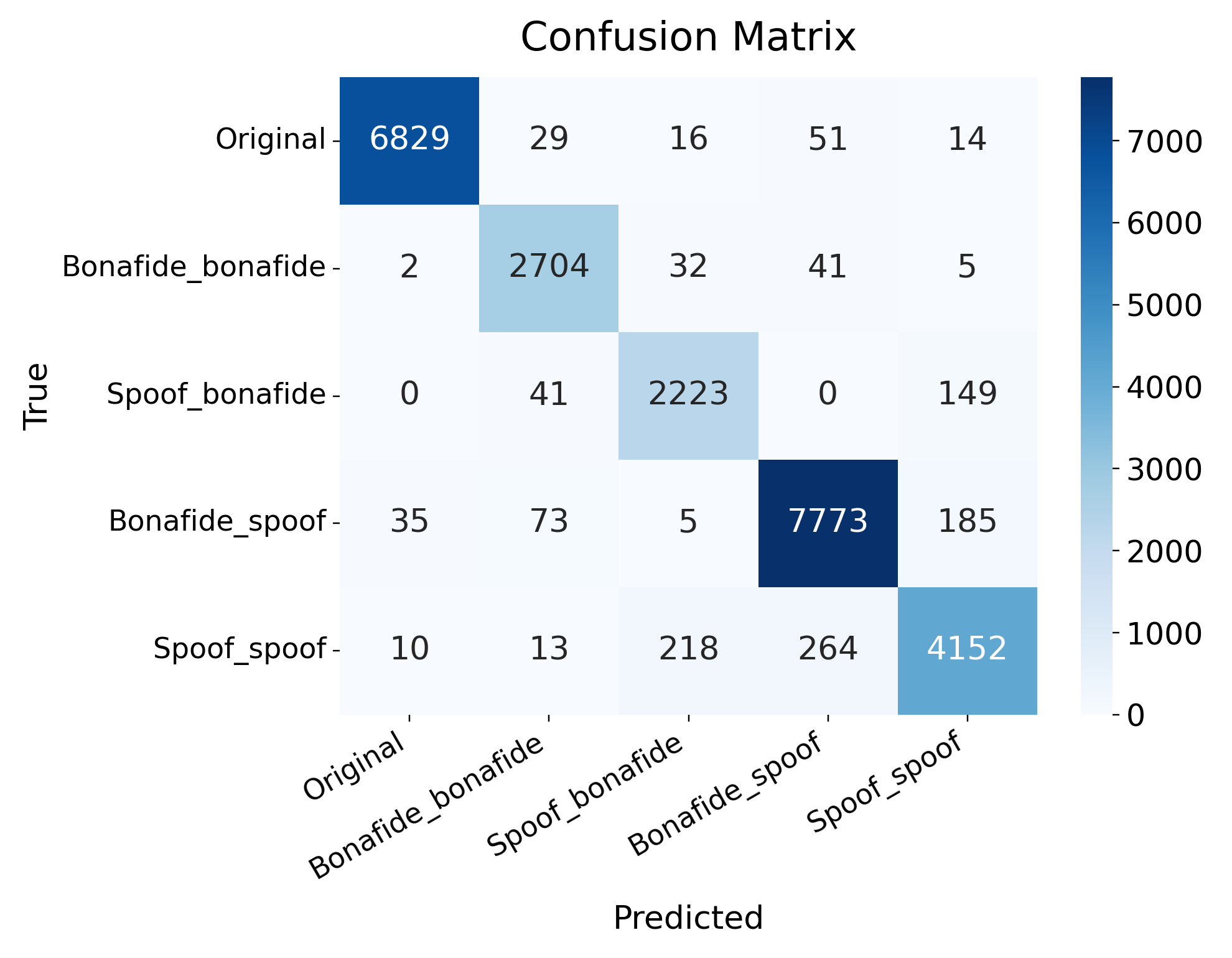}
  {\small (c)}
\end{minipage}

\caption{Training loss and validation F1-score (a), t-SNE visualization of validation representations (b), and confusion matrix on the validation set (c).}
\label{2}
\end{figure*}
\section{Results Discussion}

\subsection{Experimental Setup}

The proposed model is trained using the Adam optimizer with an initial learning rate of 1e-4, a batch size of 64, and for 12 epochs, with a learning rate scheduler applied during training. The input audio is first resampled to a fixed sampling rate. A weighted multi-task loss is adopted, where the losses for speech and environmental components are assigned a weight of 1.0, while the original component is down-weighted to 0.2 to prevent it from dominating shared representation learning. In addition, a ranking-based regularization term is introduced to enforce consistency between speech and environmental predictions under asymmetric conditions, with its weight set to 0.5. The final loss is a weighted combination of the three classification losses and the ranking loss. To improve generalization, data augmentation and oversampling are applied during training. The data augmentation includes various mixing schemes between speech and environmental sound, as well as random noise injection at different SNR levels, while oversampling is used to alleviate class imbalance. All experiments are implemented in PyTorch and conducted on a single NVIDIA RTX 4090 GPU with 24~GB of memory. The model is evaluated using multiple metrics, including classification performance (F1-score) and Equal Error Rate (EER). 

\subsection{Results and Discussion}
Experimental results in Table~\ref{tab:baseline_results} demonstrate that the proposed method consistently outperforms the baseline system in terms of overall classification performance on the CompSpoofV2 dataset. Specifically, the proposed approach achieves an F1-score of 70.11\% and 70.20\% on the evaluation and test sets, respectively, compared to 62.24\% and 63.27\% obtained by the baseline. This corresponds to an absolute improvement of approximately 7--8\%, indicating a substantial enhancement in multi-class deepfake detection capability. Additionally, Fig.~\ref{2} shows that the validation loss decreases while the validation F1-score improves and stabilizes, along with well-separated feature clusters in the t-SNE plot and strong diagonal dominance in the confusion matrix, with minor confusion between similar classes.



As shown in Table~\ref{tab3}, the proposed framework consistently outperforms single-encoder and baseline models, demonstrating the effectiveness of jointly modeling speech and environmental representations. In particular, the dual-branch design with cross-attention enables complementary feature interaction, leading to improved discrimination of component-level manipulations. The ablation results further confirm that each component, including data sampling, task-weighted loss, Matching Head, and fusion strategy, contributes to the overall performance. Compared to single-branch models (XLS-R or BEATs), the proposed method achieves better generalization by capturing both speech and environmental inconsistencies, highlighting the advantage of the proposed architecture for realistic deepfake detection scenarios.

Table~\ref{tab3} presents the computational efficiency of the proposed framework. The results show that the proposed method achieves favorable efficiency, requiring approximately 1 hour per epoch compared to about 4 hours for the baseline, while maintaining superior performance. Although it is slower than lightweight single-encoder models such as BEATs and XLS-R, the proposed approach provides a better trade-off between efficiency and performance. This indicates that integrating dual encoders with cross-attention introduces only moderate computational overhead while yielding significant improvements in detection capability.

Overall, the results validate the effectiveness of the proposed approach for environment-aware deepfake detection. The method achieves a favorable balance between detection performance and computational efficiency, demonstrating its superiority over the baseline in handling complex, component-level spoofing conditions.

\section{Conclusion}

In this work, component-level deepfake detection in the ESDD2 challenge is addressed using a dual-branch framework that jointly models speech and environmental sound through complementary self-supervised models, XLS-R and BEATs. A Matching Head is introduced to model representation differences for estimating the original class, while multi-head cross-attention enables effective interaction between speech and environmental branches. The refined representations are processed by an AASIST classifier to estimate spoofing probabilities at the component level. The proposed system achieves an approximate 7--8\% improvement in F1-score on both the evaluation and test sets compared to the baseline, along with a significant reduction in environmental EER, demonstrating the effectiveness of the proposed approach.These results demonstrate improved performance over baseline methods. Future work will focus on advanced fusion strategies and robustness under real-world conditions.

\bibliographystyle{IEEEbib}
\bibliography{ref}

\end{document}